\RequirePackage{lineno}
\documentclass[aps,prb,twocolumn,superscriptaddress,showpacs]{revtex4-1}
\usepackage{graphicx}
\usepackage{multirow}
\begin{document}
\title{Formation and upper critical fields of the two distinct A15 phases in the subelements of Powder-In-Tube \Nb~wires}
\author{Carmine Senatore}\email{Carmine.Senatore@unige.ch}
\affiliation{D\'{e}partement de Physique de la Mati\`{e}re Condens\'{e}e, Universit\'{e} de Gen\`{e}ve, Quai Ernest-Ansermet 24, 1211 Gen\`{e}ve, Switzerland}
\affiliation{D\'{e}partement de Physique Appliqu\'{e}e - Universit\'{e} de Gen\`{e}ve, rue de l'\'{E}cole de M\'{e}decine 20, 1211 Gen\`{e}ve, Switzerland}
\author{Ren\'{e} Fl\"{u}kiger}
\affiliation{D\'{e}partement de Physique de la Mati\`{e}re Condens\'{e}e, Universit\'{e} de Gen\`{e}ve, Quai Ernest-Ansermet 24, 1211 Gen\`{e}ve, Switzerland}
\affiliation{D\'{e}partement de Physique Appliqu\'{e}e - Universit\'{e} de Gen\`{e}ve, rue de l'\'{E}cole de M\'{e}decine 20, 1211 Gen\`{e}ve, Switzerland}
\newcommand{\Nb}{Nb$_{3}$Sn}
\newcommand{\NS}{Nb$_{6}$Sn$_{5}$}
\newcommand{\Bc}{B$_{c2}$}
\newcommand{\Tc}{T$_{c}$}
\newcommand{\Jc}{J$_{c}$}
\newcommand{\Tca}{$<T_{c}>$}
\begin{abstract}
It is well known that the A15 layer in the subelements of a Powder-In-Tube (PIT) \Nb~wire exhibits two different grain morphologies: a region with fine grains ($\sim$200 nm in size) representing about 60\% of the total A15 area and one with large grains (1-2 $\mu$m in size). By means of high field specific heat and magnetization measurements we have shown that these two A15 phases correspond to two distinctly different \Tc~distributions, the large grains region exhibiting a higher \Tc~and a lower \Bc, the fine grains region a lower \Tc~and a higher \Bc. We report here the values of the superconducting parameters (\Tc, \Bc) of the two A15 phases, as determined from an original model to fit the experimental \Tc~distribution. After a prolonged reaction treatment (625$^{\circ}$C/320h), an increase of the fine grain region was observed at the expenses of the large grain region, the \Bc(0K) value of the former being raised from 28.8 to 31.7 T. These changes explain the marked increase of \Jc~to 2'700 A/cm$^2$ at 4.2K/12T, the highest value measured so far in PIT wires.
\end{abstract}
 \maketitle
%
%\setpagewiselinenumbers
%\modulolinenumbers[5]
%\linenumbers
High performance Powder-In-Tube (PIT) \Nb~wires are currently being developed and tested for the next generation of accelerator magnets, in particular for the Large Hadron Collider (LHC) upgrade with high luminosity at CERN. In recent years there has been a remarkable increase of the critical current density \Jc~at the operating temperature and magnetic field. After optimization of the reaction schedule the PIT conductor developed in the frame of the EU-supported Next European Dipole (NED) project has achieved a non-Cu \Jc~value of 2'700 A/mm$^2$ at 4.2 K and 12 T \cite{Boutboul}. Further improvements require the simultaneous optimization of Sn composition and grain morphology which can be achieved through a deeper understanding of the interplay between reaction conditions and superconducting properties.
It is well known that various regions can be distinguished in the subelements of PIT wires after reaction: a layer of large ($\sim$1 $\mu$m in size) A15 grains enclosed in a layer of fine ($\sim$200 nm in size) A15 grains, all surrounded by the unreacted Nb or Nb7.5wt.\%Ta tube. The reaction process also results in a radial compositional gradient, the average Sn composition being 22 at.\% and 24.5 at.\% in the fine and large grain regions, respectively \cite{Cantoni,Hawes}. However, the precise superconducting parameters for the two regions were not precisely known. In the present work, we have determined the values of the superconducting parameters, the critical temperature \Tc~and the upper critical field \Bc, of the two A15 regions by performing low temperature specific heat measurements. This technique allows the determination of the \Tc~distribution within the A15 layer \cite{Senatore}, ruling out percolation and/or shielding effects generally present in the conventional resistive and magnetic measurements.
The examined wire is the PIT conductor B215 manufactured by ShapeMetal Innovation (SMI, now part of Bruker EAS) for the NED project. The conductor has a diameter of 1.25 mm and contains 288 Nb-7.5wt\%Ta tube subelements, whose size is $\sim$50 $\mu$m. The Sn precursor consists of a mixture of NbSn$_2$ and Sn particles, a Cu layer being interposed between these powders and the surrounding NbTa tube. Two different heat treatments were performed: the SMI-recommended reaction schedule of 84h at 675$^{\circ}$C and the optimized reaction schedule of 320h at 625$^{\circ}$C that leads to enhancement of \Jc(4.2K,12T) up to 2'700 A/mm$^2$ \cite{Boutboul}.
\begin{table*}[ht]
\caption{\label{tab1}Summary of the physical properties of the two A15 phases }
%\begin{tabular} {|c|c|c|c|c|}
\begin{tabular} {ccccc}
 \hline
%\multirow{2}{*}{} & \multicolumn{2}{|c|}{675$^{\circ}$C/84h} & \multicolumn{2}{|c|}{625$^{\circ}$C/320h} \\ \cline{2-5}
 \multirow{2}{*}{} & \multicolumn{2}{c}{675$^{\circ}$C/84h} & \multicolumn{2}{c}{625$^{\circ}$C/320h} \\ \cline{2-5}
 & Large grains & Fine grains & Large grains & Fine grains \\ \hline
 \Tca [K] & 18.2 & 18.0 & 18.35 & 17.83 \\ \hline
 $\alpha$ & 5 & 9 & 7 & 6 \\ \hline
 $\sigma$ & 0.4 & 0.9 & 0.3 & 0.83 \\ \hline
 $|dB_{c2}/dT|_{T_c}$ [T K$^{-1}$] & 1.96 & 2.35 & 2.04 & 2.58 \\ \hline
 \Bc(T=0K) [T] & 24.6 & 28.8 & 25.7 & 31.7 \\ \hline
\end{tabular}
\end{table*}
\begin{figure}
\centering \includegraphics[width=8 cm]{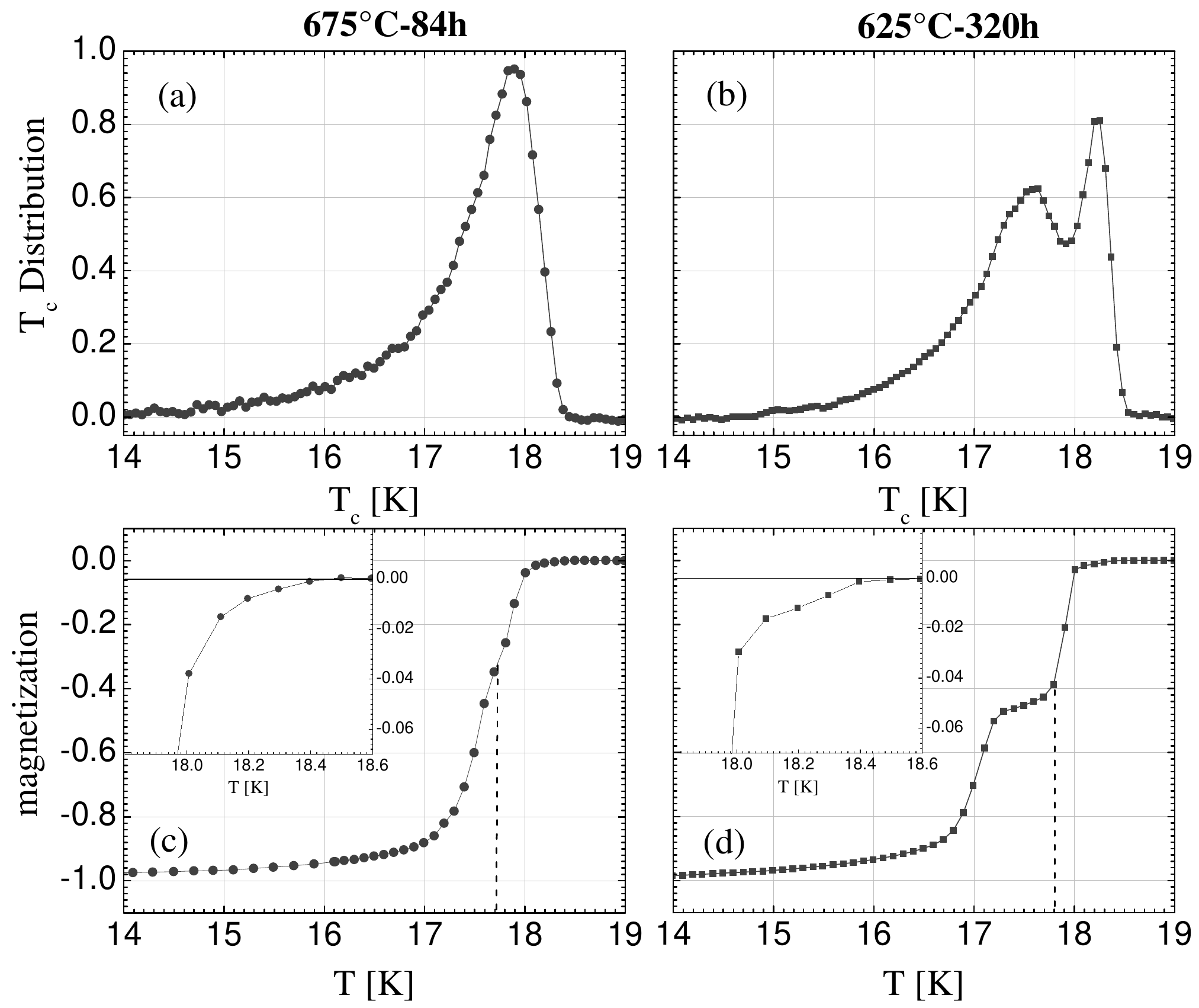} \caption{\label{fig1}
Distribution of \Tc~for the PIT wire B215, after reaction at 675$^{\circ}$C for 84h (a) and at 625$^{\circ}$C for 320h (b). The corresponding magnetization vs. temperature curves are reported in (c) and (d) and magnifications of the transition onset are included as insets. Two peaks are present in the \Tc~distribution of the wire reacted at 625$^{\circ}$C. A double superconducting transition is also present in the M(T) curves.  }
\end{figure}

The specific heat of the wires was measured from 2 to 25K at magnetic fields (B) of 0, 0.5, 1, 3, 7 and 14T, using a long relaxation technique. The distribution of \Tc~inside the filaments was determined on the superconducting transition, which was isolated from the background of phonons and normal state electrons following the procedure described in Ref. \cite{Senatore}. As a consequence of the Sn composition gradient present in the filaments the \Tc~distribution for \Nb~wires usually extends down to 14K. Moreover, the distribution curve exhibits a negative skew reflecting the shape of the \Tc~vs. at.\% Sn relation. As shown in Ref. \cite{Flukiger} and \cite{Godeke}, the variation of \Tc~with Sn content is essentially linear up to ~24 at.\% and levels off between 24 and 25 at.\%.  Therefore, the skewness of the \Tc~distribution becomes more pronounced for average Sn compositions approaching stochiometry.
Fig.~\ref{fig1} illustrates the comparison between the \Tc~distributions for the wire reacted at 675$^{\circ}$C for 84h, following the instructions of the manufacturer and at 625$^{\circ}$C for 320h, according to the optimized reaction. For both reaction schedules this wire exhibits an onset \Tc~of 18.4K. In the \Tc~distribution of the wire reacted at 625$^{\circ}$C for 320h two distinct peaks were observed. This can be related to the morphology of the A15 phases. The two different grain morphologies are the consequence of the complex A15 formation process. As reaction heat treatment starts, several Cu-Sn intermetallic phases are formed from precursors and NbSn$_2$ powders dissolution is almost complete below 440$^{\circ}$C \cite{DiMichiel}. At about 550$^{\circ}$C, the first reaction occurring in the Nb tubes is the reformation of NbSn$_2$\cite{DiMichiel}. Subsequently Sn leaves NbSn$_2$ and \NS~large grains grow inwards the Nb tube. A first A15 phase is formed on the outside of the \NS~layer, mainly due to the Sn diffusion from the core. As the reaction proceeds, the \NS~phase is converted into \Nb, too. However the grains of this second \Nb~phase retain the size of the \NS~grains, which is significantly larger than that of the first A15 layer directly formed from the solid state diffusion of Sn into Nb.
\begin{figure}
\centering \includegraphics[width=6 cm]{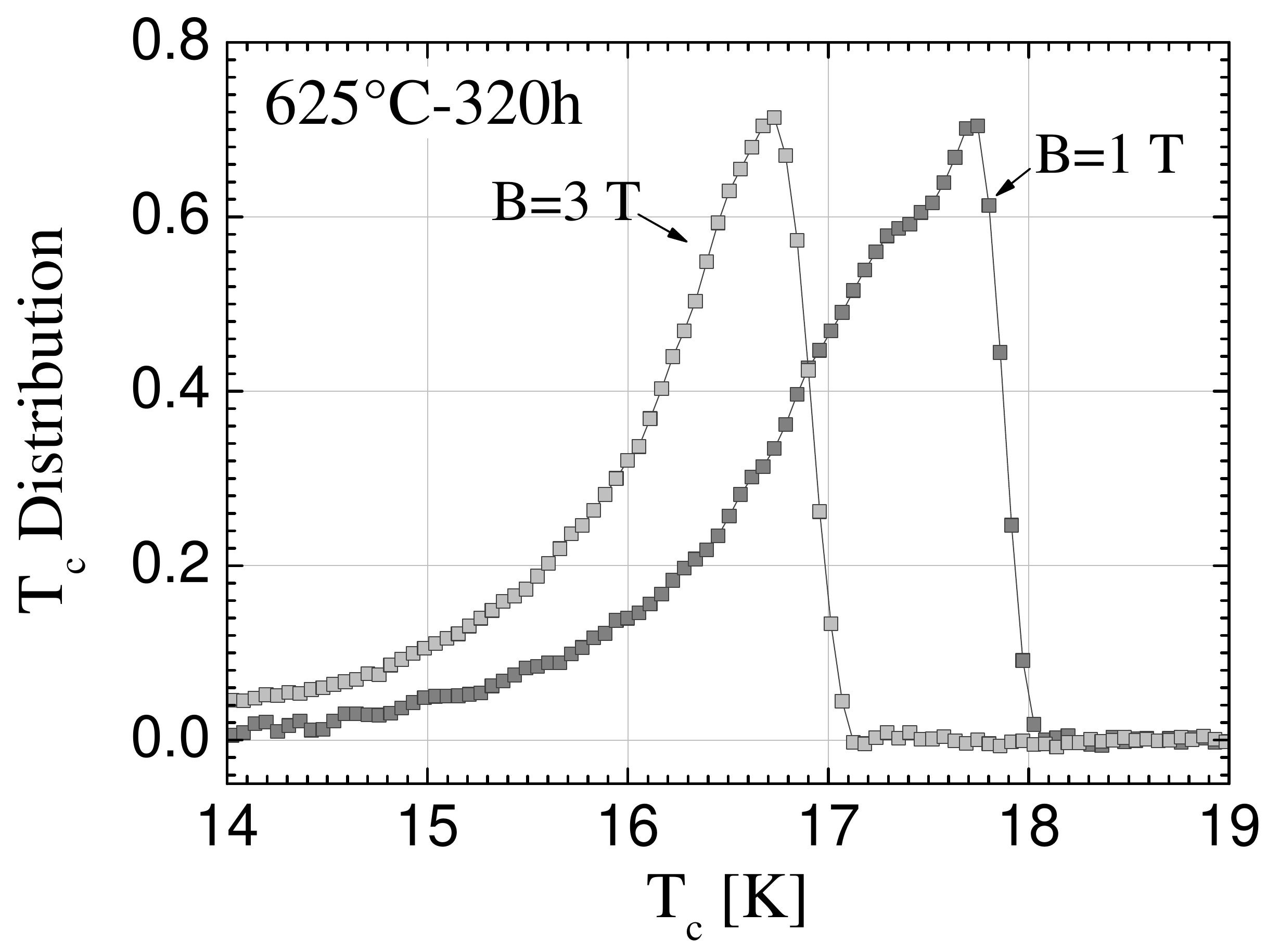} \caption{\label{fig2}
\Tc~distribution at B = 1T and B = 3T for the wire reacted at 625$^{\circ}$C for 320h. The two peaks tend to merge as the magnetic field is increased.}
\end{figure}
\begin{figure}
\centering \includegraphics[width=6 cm]{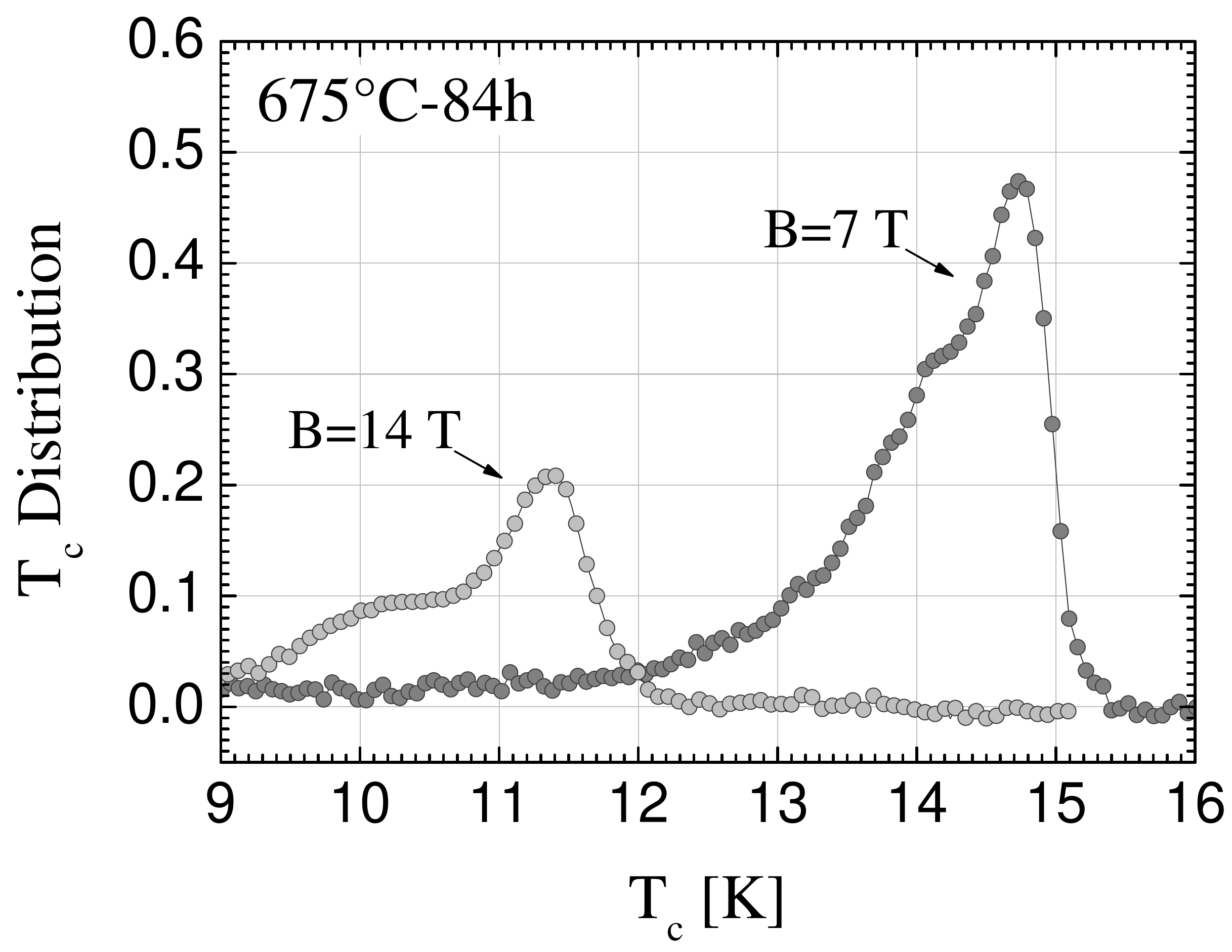} \caption{\label{fig3}
\Tc~distribution at B = 7T and B = 14T for the wire reacted at 675$^{\circ}$C for 84h. The contribution of large grains appears as a shoulder below the peak temperature.}
\end{figure}
The two peaks in the \Tc~distribution thus reflect the relative volume of the large and fine grain regions. As \Tc~depends on the Sn content, the peak with maximum at 18.2K corresponds to the large grains, whose composition is close to the stoichiometry(average 24.5 at\% Sn) as revealed from the Energy Dispersive X-ray Spectroscopy (EDS) analysis reported in Ref.\cite{Cantoni}. The \Tc~distribution with a maximum at 17.6K is correlated to the fine grains, whose average content is $\sim$22 at.\% \cite{Cantoni}. The presence of a bimodal \Tc~distribution was recently observed in \Nb~bulk samples prepared with the addition of Cu \cite{Mentik}. From the comparison with the results obtained on binary \Nb~samples, the authors conclude that Cu addition triggers the formation of the off-stoichiometric A15 phase responsible for the low-\Tc~peak in the \Tc~distribution.
In the present work, the superconducting transition was also characterized using a SQUID magnetometer. The magnetization vs. temperature curve, measured at 1 mT and reported in Fig.~\ref{fig1}d, also shows two superconducting transitions, thus confirming the specific heat measurements. The lower \Tc~transition is detected because the higher \Tc~part is on the inside of the filament, and thus there are no shielding effects. This information about the spatial variation of \Tc~confirms the identification of the two peaks in the \Tc~distribution with the two observed grain regions. We also observe the contributions from the two grain families in the inductive \Tc~measurement performed on the PIT wire reacted at 675$^{\circ}$C for 84h (Fig.~\ref{fig1}c).
However, the onset \Tc~from the \Tc~distribution is higher by $\sim$0.1K compared to the value determined from magnetization , due to the higher sensitivity of the calorimetric technique. In fact, calorimetry detects the thermodynamic transition, whereas magnetometry reveals the presence of screening currents.
\begin{figure}
\centering \includegraphics[width=6 cm]{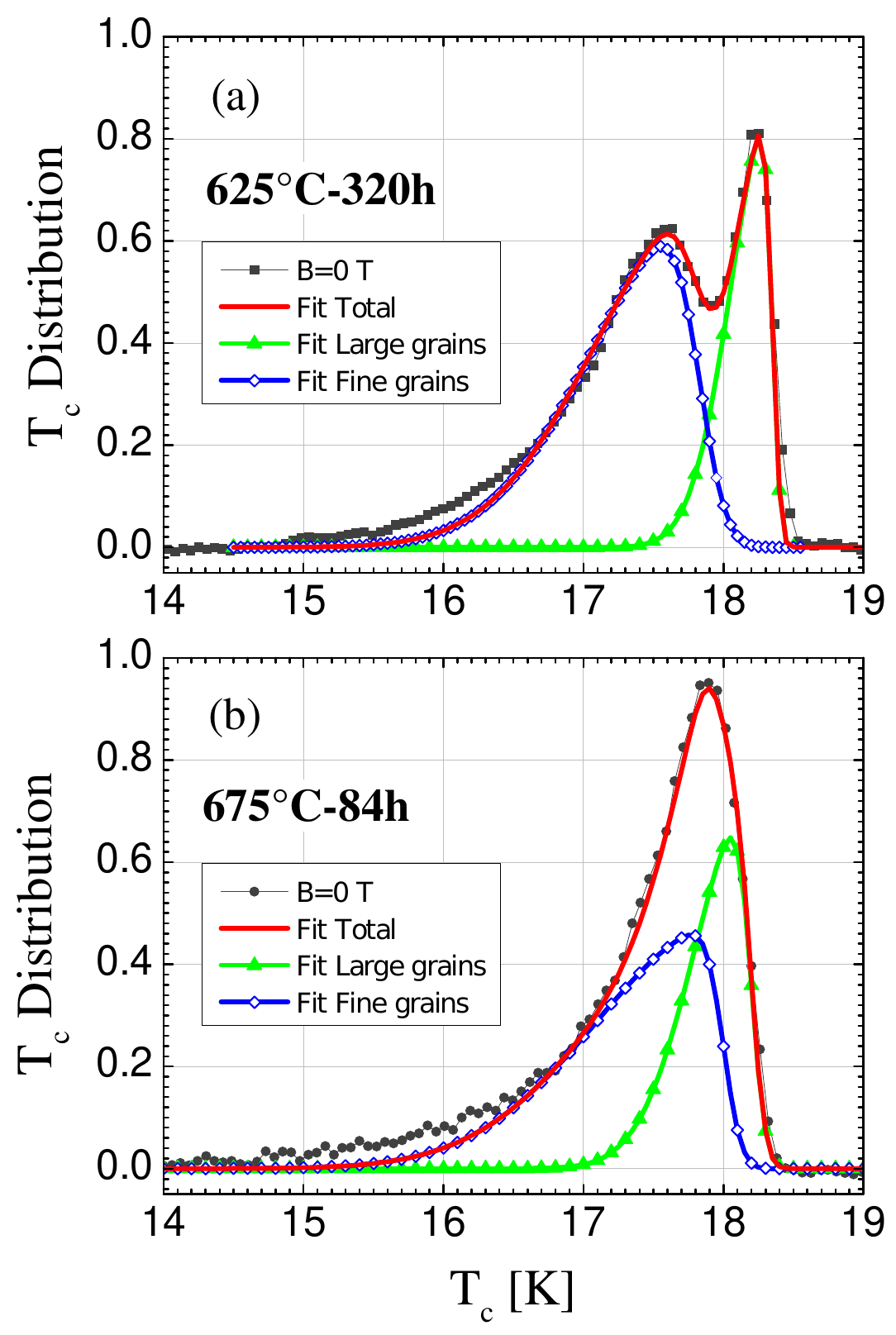} \caption{\label{fig4}
Fit of the \Tc~distribution at B = 0T for the wires reacted at 625$^{\circ}$C for 320h (a) and 675$^{\circ}$C for 84h (b). The contributions from large grains (solid triangles) and fine grains (open diamonds) are reported.}
\end{figure}

As an increasing magnetic field is applied, the two peaks in the \Tc~distribution of the wire reacted at 625$^{\circ}$C for 320h shift towards lower temperatures, but at different rates (Fig.~\ref{fig2}), reflecting the different \Bc~of the two grain families. In particular, we deduce that large grains have lower \Bc~values. In fact, the field induced shift of the peak corresponding to large grains is more pronounced compared to the behaviour of the peak associated to fine grains. As a consequence the two peaks tend to merge for increasing magnetic fields, as shown in Fig.~\ref{fig2}. At B = 3T the \Tc~distribution exhibits a single peak. The reduced \Bc~in the large grain region is a consequence of the higher average Sn content: this leads to a higher atomic ordering and, thus, to a lower normal state resistivity \cite{Flukiger}. For the wire reacted at 675$^{\circ}$C for 84h, it is not possible to discern the two contributions in the \Tc~distribution at B = 0T, thus indicating a large overlap between the Sn composition ranges of the two grain families. However, a difference in the \Bc~values is still present, even if less pronounced compared to the wire reacted at 625$^{\circ}$C. As the applied magnetic field is large enough, the contribution to the \Tc~distribution of large grains, which have a lower \Bc, appears as a shoulder in the curve at low temperature. This is shown in Fig.~\ref{fig3} for B = 7T and 14T.
\begin{figure}
\centering \includegraphics[width=6 cm]{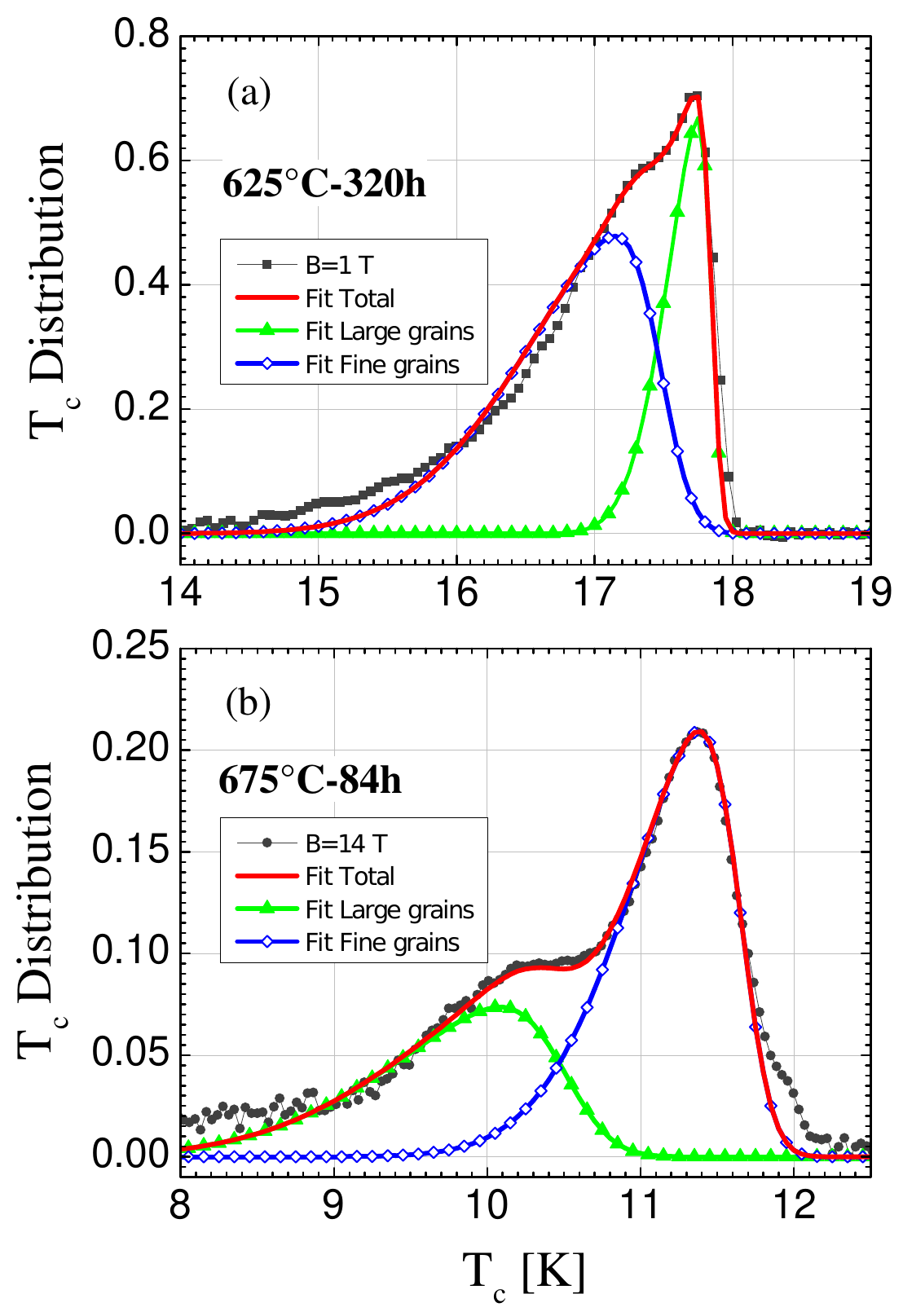} \caption{\label{fig5}
Fit of the \Tc~distribution measured in the presence of a magnetic field: B = 1T for the wire reacted at 625$^{\circ}$C for 320h (a) and B = 14T for the wire reacted at 675$^{\circ}$C for 84h (b). The contributions from large grains (solid triangles) and fine grains (open diamonds) are reported.}
\end{figure}
The \Tc~distribution for these PIT wires is thus the superposition of two distribution curves resulting from the two distinct A15 regions. In order to determine the average values of \Tc~and \Bc~for fine and large grains, we developed a model to fit the \Tc~distribution at various magnetic fields. We describe the contribution of each A15 phase as an independent skew normal distribution. The skew normal distribution is a probability function that generalizes the Gaussian distribution to account for the skewness arising from the non-linearity in the \Tc~vs. at.\% Sn relation. It is defined by three parameters: the expected value for \Tc, \Tca, the variance  $\sigma$, reflecting the width of the Sn composition gradient, and the skewness  $\alpha$, whose absolute value increases approaching stochiometry in the Sn composition. The probability density function is given by \cite{Azzalini}:
\begin{equation}\label{fit1}
    f(T) = \frac{2}{\sigma}\phi\Bigr(\frac{T-<T_{c}>}{\sigma}\Bigl)\Phi\Bigr(\alpha\frac{T-<T_{c}>}{\sigma}\Bigl) \, .
\end{equation}
$\phi(x)$ denotes the standard normal distribution
\begin{equation}\label{fit2}
    \phi(x) = \frac{1}{\sqrt{2\pi}}e^{-\frac{x^2}{2}} \, ,
\end{equation}
with the cumulative distribution function given by
\begin{equation}\label{fit3}
    \Phi(x) = \int_{-\infty}^x \! \phi(t) \, \mathrm{d} t.  \,
\end{equation}
Fig.~\ref{fig4} shows the fits of the \Tc~distribution for the wires reacted at 625$^{\circ}$C for 320h (a) and at 675$^{\circ}$C for 84h (b). We also report in this figure the two contributions to the \Tc~distribution from large grains (solid triangles) and fine grains (open diamonds). The volume fraction of the A15 layer occupied by each of the two grain families is determined by integrating the corresponding contribution to the \Tc~distribution over the explored range of temperatures. It follows that after standard reaction fine grains represent ~60\% of the A15 in the wire. This value is raised to ~65\% in the sample reacted according to the optimized schedule. The  ratio of fine grains is consistent with the estimation of Boutboul et al.\cite{Boutboul} on the basis of micrographs on fractured samples. Their Scanning Electron Microscope (SEM) examination showed that $\lesssim$70\% of the A15 area is composed by fine grains, for both reaction schedules.
From the analysis of the fit results, summarized in table~\ref{tab1}, it follows that the reaction heat treatment at lower temperature for longer duration leads to the increase of \Tca~in the large grain region (\Tca~= 18.35K, compared to 18.2K after standard reaction) and to the decrease of \Tca~in the fine grain region (\Tca~= 17.83K, compared to 18K after standard reaction). The reaction parameters also have an influence on the skewness: after heat treatment at 625$^{\circ}$C for 320h the skewness in the component of the \Tc~distribution corresponding to the large grains is increased ($\alpha$=7, compared to  $\alpha$=5 after standard reaction). For fine grains the skewness of the \Tc~distribution becomes less pronounced ($\alpha$=6, compared to  $\alpha$=9 after standard reaction). These results reflect the effect of the optimized heat treatment at 625$^{\circ}$C on the Sn composition of the two A15 regions. In particular we argue that in the large grain region the average Sn content is increased while it is decreased in the fine grain region. On the other hand, we observe that the  $\sigma$ values of both large and fine grains do not depend significantly on the heat treatment parameters. This implies that reaction conditions have an influence on the average Sn content in each of the two grain regions but not on the width of the Sn composition gradient.
\begin{figure}
\centering \includegraphics[width=6 cm]{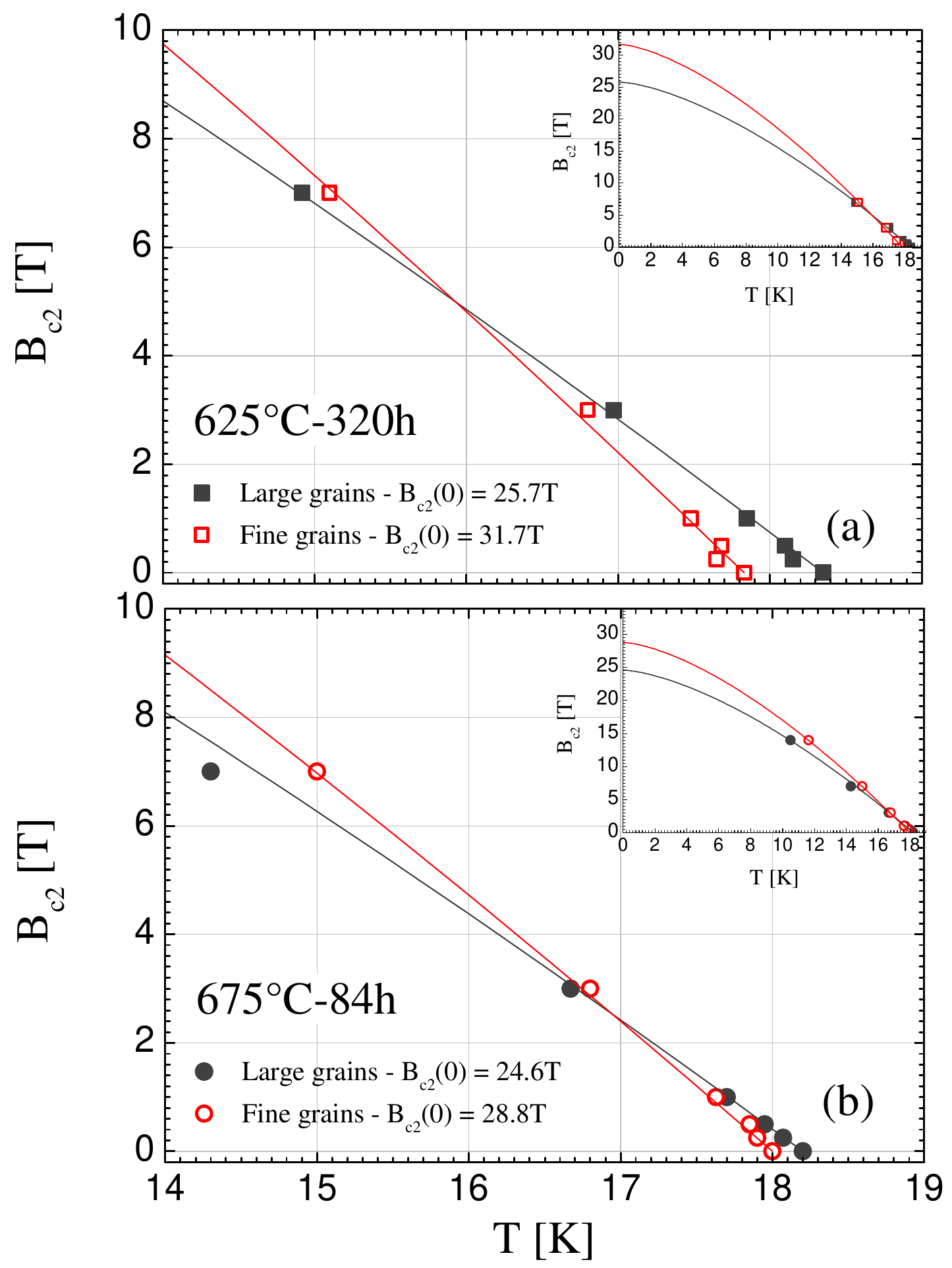}
\caption{\label{fig6}
\Bc~plot for the two A15 phases, as determined from the fit of the \Tc~distribution measured at various magnetic fields. (a) Sample reacted at 625$^{\circ}$C; (b) sample reacted at 675$^{\circ}$C. In the inset the \Bc~values are extrapolated at T = 0K, following the WHH fitting model. The optimized heat treatment at 625$^{\circ}$C leads to a large improvement of \Bc~in the fine grains. }
\end{figure}
Information regarding the upper critical field \Bc~of the two A15 regions were obtained extending the fit analysis to the \Tc~distribution measured in the presence of an external magnetic field. Fig.~\ref{fig5} shows two examples of fitting: the \Tc~distribution at B = 1T for the wire reacted at 625$^{\circ}$C for 320h (a) and at B = 14T for the wire reacted at 675$^{\circ}$C for 84h (b). We were able to determine the \Tca(B) - or, equivalently, \Bc(T) - dependence for both the A15 regions in a wide range of magnetic fields (Fig.~\ref{fig6}). The \Bc(0K) values were determined from the data up to 7T, i.e. in the range where the linear approximation of the \Bc(T) dependence holds, based on the standard Werthamer-Helfand-Hohenberg (WHH) fitting formula\cite{Werthamer}:
\begin{equation}\label{Bc2}
    B_{c2}(0K) = 0.693 <T_c> \Bigr\vert\frac{\mathrm{d}B_{c2}}{\mathrm{d}T}\Bigl\vert_{T_c} \, .
\end{equation}
Our results are reported in table I, indicating that the optimized heat treatment influences the growth of both large and fine grains, but with a much stronger increase of \Bc~for the fine grains, where a shift from 28.8T to 31.7T was observed, compared to an improvement for large grains from 24.6T to 25.7T. The extrapolated values of \Bc(0K)=31.7T for the fine grain region in the wire reacted at 625$^{\circ}$C is rather high. On the other hand, in Ref.\cite{Godeke2} Godeke et al. extract from resistivity measurements up to 30T a \Bc(0K) of $\sim$29T for a PIT wire reacted at 675$^{\circ}$C for 64h, thus in agreement with the value determined by us for the fine grains after standard reaction.

In summary, we have investigated the differences in \Tc~and \Bc~for the two A15 phases present in the superconducting layer of PIT \Nb~wires by high field calorimetry. After two largely different heat treatments we found in both cases that higher \Bc~values are achieved in fine grains, in spite of the lower \Tc, as a consequence of the enhancement of normal state electrical resistivity for lower average Sn contents. The differences in \Tc~and \Bc~between the two A15 phases with different grain sizes are more pronounced for the wire sample reacted at lower temperature for longer duration. The prolonged heat treatmentat 625$^{\circ}$C facilitates the formation of an A15 phase with inferior \Tc~ and Sn content but superior \Bc~at 4.2K.  This reflects the influence of reaction kinetics on the atomic ordering and, thus, on the final Sn composition. Since fine grains are directly formed from the Sn diffusion into Nb, while large grains from the Sn depletion of the \NS~phase, we suggest that the two A15 regions follow different ordering kinetics. In particular, measurements performed on samples reacted for up to 500h show that thermodynamic equilibrium is never achieved even at the end of the heat treatment; we argue that the two grain families keep a different ordering with the fine grains characterized by a higher disorder and, thus, higher \Bc~values.

This work was supported by the Swiss National Science Foundation through the National Centre of Competence in Research - Materials with Novel Electronic Properties (MaNEP/NCCR). A. Ubaldini, B. Seeber, M. Bonura, G. Mondonico and D. Zurmuehle provided valuable assistance.

\end{document}